\documentstyle[12pt]{article}
\def\hybrid{\topmargin -20pt	\oddsidemargin 0pt
	\headheight 0pt	\headsep 0pt
	\textwidth 6.25in	
	\textheight 9.5in	
	\marginparwidth .875in
	\parskip 5pt plus 1pt	\jot = 1.5ex}

\hybrid

\def\baselinestretch{1.2}

\catcode`\@=11

\def\marginnote#1{}
\newcount\hour
\newcount\minute
\newtoks\amorpm
\hour=\time\divide\hour by60
\minute=\time{\multiply\hour by60 \global\advance\minute by-\hour}
\edef\standardtime{{\ifnum\hour<12 \global\amorpm={am}%
	\else\global\amorpm={pm}\advance\hour by-12 \fi
	\ifnum\hour=0 \hour=12 \fi
	\number\hour:\ifnum\minute<10 0\fi\number\minute\the\amorpm}}
\edef\militarytime{\number\hour:\ifnum\minute<10 0\fi\number\minute}

\def\draftlabel#1{{\@bsphack\if@filesw {\let\thepage\relax
   \xdef\@gtempa{\write\@auxout{\string
      \newlabel{#1}{{\@currentlabel}{\thepage}}}}}\@gtempa
   \if@nobreak \ifvmode\nobreak\fi\fi\fi\@esphack}
	\gdef\@eqnlabel{#1}}
\def\@eqnlabel{}
\def\@vacuum{}
\def\draftmarginnote#1{\marginpar{\raggedright\scriptsize\tt#1}}

\def\draft{\oddsidemargin -.5truein
	\def\@oddfoot{\sl preliminary draft \hfil
	\rm\thepage\hfil\sl\today\quad\militarytime}
	\let\@evenfoot\@oddfoot	\overfullrule 3pt
	\let\label=\draftlabel
	\let\marginnote=\draftmarginnote
   \def\@eqnnum{(\theequation)\rlap{\kern\marginparsep\tt\@eqnlabel}%
\global\let\@eqnlabel\@vacuum}  }

\def\preprint{\twocolumn\sloppy\flushbottom\parindent 2em
	\leftmargini 2em\leftmarginv .5em\leftmarginvi .5em
	\oddsidemargin -.5in	\evensidemargin -.5in
	\columnsep .4in	\footheight 0pt
	\textwidth 10.in	\topmargin  -.4in
	\headheight 12pt \topskip .4in
	\textheight 6.9in \footskip 0pt
	\def\@oddhead{\thepage\hfil\addtocounter{page}{1}\thepage}
	\let\@evenhead\@oddhead	\def\@oddfoot{}	\def\@evenfoot{} }

\def\numberbysection{\@addtoreset{equation}{section}
	\def\theequation{\thesection.\arabic{equation}}}

\def\underline#1{\relax\ifmmode\@@underline#1\else
	$\@@underline{\hbox{#1}}$\relax\fi}

\def\titlepage{\@restonecolfalse\if@twocolumn\@restonecoltrue\onecolumn
     \else \newpage \fi \thispagestyle{empty}\c@page\z@
	\def\thefootnote{\fnsymbol{footnote}} }

\def\endtitlepage{\if@restonecol\twocolumn \else \newpage \fi
	\def\thefootnote{\arabic{footnote}}
	\setcounter{footnote}{0}}  

\catcode`@=12
\relax

\def\figcap{\section*{Figure Captions\markboth
	{FIGURECAPTIONS}{FIGURECAPTIONS}}\list
	{Figure \arabic{enumi}:\hfill}{\settowidth\labelwidth{Figure 999:}
	\leftmargin\labelwidth
	\advance\leftmargin\labelsep\usecounter{enumi}}}
 \relax
\def\tablecap{\section*{Table Captions\markboth
	{TABLECAPTIONS}{TABLECAPTIONS}}\list
	{Table \arabic{enumi}:\hfill}{\settowidth\labelwidth{Table 999:}
	\leftmargin\labelwidth
	\advance\leftmargin\labelsep\usecounter{enumi}}}
 \relax
\def\reflist{\section*{References\markboth
	{REFLIST}{REFLIST}}\list
	{[\arabic{enumi}]\hfill}{\settowidth\labelwidth{[999]}
	\leftmargin\labelwidth
	\advance\leftmargin\labelsep\usecounter{enumi}}}
 \relax
\makeatletter
\newcounter{pubctr}
\def\publist{\@ifnextchar[{\@publist}{\@@publist}}
\def\@publist[#1]{\list
	{[\arabic{pubctr}]\hfill}{\settowidth\labelwidth{[999]}
	\leftmargin\labelwidth
	\advance\leftmargin\labelsep
	\@nmbrlisttrue\def\@listctr{pubctr}
	\setcounter{pubctr}{#1}\addtocounter{pubctr}{-1}}}
\def\@@publist{\list
	{[\arabic{pubctr}]\hfill}{\settowidth\labelwidth{[999]}
	\leftmargin\labelwidth
	\advance\leftmargin\labelsep
	\@nmbrlisttrue\def\@listctr{pubctr}}}
 \relax
\makeatother
\newskip\humongous \humongous=0pt plus 1000pt minus 1000pt
\def\caja{\mathsurround=0pt}
\def\eqalign#1{\,\vcenter{\openup1\jot \caja
	\ialign{\strut \hfil$\displaystyle{##}$&$
	\displaystyle{{}##}$\hfil\crcr#1\crcr}}\,}
\newif\ifdtup

\relax

\def\thefootnote{\fnsymbol{footnote}}
\def\be{\begin{equation}}
\def\ee{\end{equation}}
\def\ba{\begin{eqnarray}}
\def\ea{\end{eqnarray}}
\def\d{\partial}

\def\s{\sigma}

\def\P{\Phi}

\def\ub{{\bar u}}
\begin{document}
\renewcommand{\theequation}{\thesection.\arabic{equation}}
\begin{titlepage}
\begin{center}

\hfill HUB-IEP-93/7\\

\vskip .2in

{\large \bf
Gravitational and Axionic Backgrounds for
Four-dimensional Superstrings\footnote{Talg given at the EPS 93 Conference,
held at Marseilles, July 22-27. To appear in these Proceedings}}
\vskip .5in

{\bf D. L\"ust}

\vskip .1in
{\em Humboldt Universit\"at zu Berlin\\
Fachbereich Physik\\
D-10099 Berlin, GERMANY}

\end{center}

\vskip 1.7in

\begin{center} {\bf ABSTRACT } \end{center}
\begin{quotation}\noindent
We construct new four-dimensional superstring vacua with extended
superconformal symmetries. A non-trivial dilaton background implies the
existence of Abelian killing symmetries. These are used to construct
dual equivalent backgrounds in a way preserving the $N=2$ superconformal
invariance.

\end{quotation}
\vskip3.0cm
December 1993\\
\end{titlepage}
\vfill
\eject
\def\baselinestretch{1.2}
\baselineskip 16 pt
\noindent

\vskip 4mm
\section{Introduction}
\vskip 2.0mm

\setcounter{equation}{0}

During recent years, a lot of efforts were focussed on  the construction
of $N=2$
superconformal  (heterotic) string backgrounds \cite{4d}.
These backgrounds lead to target space supersymmetry and, consequently,
 the perturbative
vacuum is guaranteed to be stable. In these works the discussion
mainly concentrated on flat four-dimensional Minkowski space-time
times an internal compact space without torsion
and with constant dilaton field, ie. tori, orbifolds
and numerous Calabi-Yau
spaces. However
we like to construct supersymmetric string vacua with, in addition
to the metric background, more general non-constant bachground fields.
Moreover,
to address certain important questions in quantum gravity one has to
consider string backgrounds which describe four-dimensional curved
space-times. In particular one is interested to construct exact
superconformal field theories which correspond to four-dimen\-sio\-nal
black-hole backgounds,  cosmological scenarios or
supersymmetric instanton type of solutions.

In this contribution  we will report about a relatively systematic
discussion \cite{kkl}
on
supersymmetric string backgrounds with $N=2$ or $N=4$ superconformal
symmetry,
based on compact as well as
non-compact spaces plus
non-trivial antisymmetric tensor-field
and non-con\-stant dilaton.
In contrast to the compact Calabi-Yau spaces, almost all backgrounds
with non-trivial dilaton field will possess Killing symmetries.
Many of such backgrounds exhibit singularities on some hypersurface
in spacetime and
 can be regarded as
generalizations of the two-dimensional black-hole considered
in \cite{bh}.

A key to the proper understanding of string propagation
on curved spaces is provided by duality symmetries \cite{duality}.
Duality symmetries relate  different backgrounds which
nevertheless
correspond to the same superconformal field theory.
 Specifically, we will show that some interesting
non-K\"ahlerian $ N=4$ solutions, which describe four-dimensional
axionic instantons, are dual-equivalent to
four-dimensional, non-compact
Ricci-flat K\"ahler spaces.

\renewcommand{\theequation}{\thesection.\arabic{equation}}

\section{The $N=2$ ($N=4$) Background and $U(1)$ Duality
Transformations}

\setcounter{equation}{0}
The most general $N=2$ superspace action
for $m$ chiral superfields $U_i$ ($i=1,\dots , m$) ($\bar D_\pm U_i =0$) and
$n$ twisted chiral superfields $V_p$ ($p=1,\dots ,n$) ($\bar D_+ V_p=D_-
V_p=0$)
in two dimensions is determined by a single real function
$K(U_i,\bar U_i,V_p,\bar V_p)$ \cite{GHR}:
\be
S={1\over 2\pi \alpha '}\int{\rm d}^2xD_+D_-\bar D_+\bar D_-
K(U_i,\bar U_i,V_p,\bar V_p).\label{action}
\ee
To see the background interpretation of the theory it is convenient
to write down the purely bosonic part of the superspace action
(\ref{action}):
\be
\eqalign{
S=&-{1\over 2\pi\alpha '
}\int{\rm d}^2x\lbrack K_{u_i\bar u_j}\partial^a u_i
\partial_a\bar u_j-K_{v_p\bar v_q}\partial^a v_p\partial_a\bar v_{ q}
\cr& +\epsilon_{ab}(K_{u_i\bar v_p}\partial_a u_i\partial_b\bar v_{
p}
+K_{v_p\bar u_i}\partial_a v_p\partial_b\bar u_{i})\rbrack,\label{osonic}\cr}
\ee
where
$K_{u_i\bar u_j}={\partial^2K\over\partial U_i\partial\bar
U_j}$, etc.
Here $u_{i}$ is the lowest component of the superfield $U_{i}$ and so
on.
Thus, one recognizes that the first two terms in above equation
describe the  in general non-K\"ahlerian metric background of the model
(the metric is K\"ahler only when $m=0$ or $n=0$). The
$\epsilon_{ab}$-term in (\ref{osonic}) provides the
antisymmetric
tensor field background. Of course, in order that these backgrounds provide
consistent string solutions, they have to satisfy the string equation
of motion, i.e. the vanishing of the $\beta$-function equations \cite{beta}.
Including also the dilaton background $\Phi(u_i,v_p)$,
the $\beta$-function equations will lead to some differential equations
for the two functions $K$ and $\Phi$ as we will discuss in the following.
 Moreover, the central charge defect $\delta c=c-{3D\over 2}$ will be
 determined by $K$ and $\Phi$. In the presence of $N=4$ superconformal
 symmetry the soloutions to lowest order in $\alpha'$ are exact
 to all orders in a specific scheme, and $\delta c$ remains ero to all orders.

Now we consider the   simple case of a single $U(1)$ isometry
assuming that the potential $K$ has one Killing symmetry, $R=Z+\bar Z$:
\be
K=K(Z+\bar Z,Y_i,\bar Y_i,V_p,\bar V_p)\label{killing}
\ee
where $Z$ and $Y_i$ are chiral fields, whereas $V_p$ are twisted
chiral
fields. (Of course the discussion holds in the same way if
$Z$ is a twisted chiral field.)
 Then
the `dual' potential has the following form \cite{LR,GHR}
\be
\tilde K(R,Y_i,\bar Y_i,V_p,\bar V_p,\Psi+\bar\Psi)=
K-R(\Psi+\bar\Psi),\label{dualpot}
\ee
where $\Psi$ a twisted chiral field.
Varying the action with respect to $\Psi$ gives back the original
theory. On the other hand one can equally well consider
the constraint coming from the variation with respect to $Z$,
\cite{LR}
$
{\delta S\over\delta Z}=0$,
and
the dual theory is obtained as a Legendre
transform of $K$ where the independent variable are $\psi$, $y_i$
and $v_p$.

\section{K\"ahler Spaces without Torsion and their Duals}

\setcounter{equation}{0}

If the antisymmetric tensor field vanishes the space is K\"ahlerian and
the metric is given in terms of the K\"ahler potential by the
standard formula
$
G_{ij}=G_{{\bar i}{\bar j}}=0$, $ G_{i{\bar
j}}=K_{u_{i}\ub_{j}}$. Then the Ricci-tensor takes its well-known form
$
R_{u_i\bar u_j}=-\partial_{u_i}\partial_{\bar u_j}
U$, $
R_{\bar u_i\bar u_j}=0$
 with $U=\log\det K_{u_i\bar u_j}={1\over 2}\log\det G$.
The only condition for conformal invariance is $\beta_{\mu\nu}^G=0$
which here implies
\be
\Phi={1\over 2}U+f(u_i)+\bar f(\bar u_i),\label{kahl}
\ee
and
\be
\nabla_{u_i}\partial_{u_j}\Phi=\nabla_{\bar u_i}\partial_{\bar u_j}
\Phi=0.\label{phikah}
\ee
where $f$ is an arbitrary holomorphic function.
As described in \cite{kkl} it is not difficult to show that the vanishing of
the
 holomorphic double derivative on the dilaton implies,
 for non-trivial dilaton, that
there is
a generic Killing symmetry in the K\"ahler metric as well as in the
dilaton. Then, in a special coordinate system
the compatibility of the equations (\ref{kahl}) and
(\ref{phikah})
along with our
freedom to perform K\"ahler transformations implies that
\be
K=K(z+\bar z,y_{i},\bar y_{i})\;\;,\;\;\P=\d_{z}K=\d_{\bar z}
K\label{fe}
\ee
and
\be
\P={1\over 2}U+C(z+\bar z)\label{ff}
\ee
where $C$ is any real number.
We can take (\ref{ff}) as the equation specifying the dilaton in
terms of the
metric and then (\ref{ff}) becomes a non-linear differential equation
for the
K\"ahler potential
\be\det [K_{u_{i}\ub_{j}}]=\exp
[-2C(z+\bar z)+K_{z}+K_{\bar z}]
\label{fg}
\ee
generalizing the CY condition.

Let us consider a special class
 assuming that the model
has a $U(N)$ isometry, i.e.
$
K=K(x)$, $\Phi=\Phi(x)$ with
$x=\sum_{i=1}^N|u_i|^2$.
For $N>1$, the linear term in the dilaton, eq.(\ref{ff}),
is not allowed by the $U(N)$ isometry and the dilaton field becomes
$
\P={1\over 2}U={1\over 2}\log
\lbrack(K')^{N-1}(K'+xK'')\rbrack$ ($K'={\partial K\over
\partial x}$).
Let us define the following function:
$
Y(x)=xK'(x)$.
Now we have to insert this ansatz
into the field equation, and the solution
 takes the following form:
\be
e^{Y}\sum_{m=0}^{N-1}{(-1)^mY^m\over m!}=A+Bx^N.\label{solution}
\ee
Here $A$ and $B$ are arbitrary parameters.The
 dilaton can be also expressed
entirely of $Y$ as
$
\P={1\over 2}U=-{1\over 2}Y+{\rm const}$.

Let us now consider four-dimensional backgrounds, namely
we consider
solutions of the form eq. (\ref{solution}) with
$N=2$.
Using $Y$ together with the overall phase $\theta$ as (real) coordinates,
the metric then reads:
\be
\eqalign{{\rm d}s^2&={({\rm d}Y)^2\over  4f(Y)}+{f(Y)\over 4}
\biggl({\rm d}\theta
-i{\bar y{\rm d} y-y{\rm d}\bar y\over 1+y\bar y}\biggr)^2\cr &
+{Y\over (1+
y\bar y)^2}
{\rm d}y{\rm d}\bar y,\quad
f(Y)={2(Ae^{-Y}+Y-1)\over Y}.\label{betterme}\cr}
\ee
This metric  in the ($\theta,\psi,\phi$)
subspaces is a deformation of the fibration of $S^{3}$ over
$S^2$, whose line element is manifest in (\ref{betterme}).
 The space possesses a generic singularity for
$Y\rightarrow\infty$, and another singularity at $Y=0$, if $A\neq 1$.

Applying eq.\ref{dualpot}, the dual metric is given
 as
\be
{\rm d}\tilde s^2={
{\rm d}\psi{\rm d}\bar\psi\over f(Y)}
+{Y\over (1+y\bar y)^2}{\rm d}y{\rm d}
\bar y,
\label{dualntwo}
\ee
whereas the dual dilaton and antisymmetric tensor field look like
$
\tilde\Phi=-{1\over 2}\log[e^Yf(Y)]$, $\tilde B_{\psi
\bar y}=2y/(1+y\bar y)$.
The dual space has
curvature singularities at $Y=-\infty,0$ and, for
generic values
of $A$ at the zeros of $f(Y)$.

\section{Four-dimensional Non-K\"ahlerian Spaces with Torsion and their Duals}

\setcounter{equation}{0}

To start with non-vanising antisymmetric tensor fields we restrict
ourselves to the simplest case, namely four-dimensional
target spaces, i.e. $m=n=1$. A particular simple class of solutions
of the $\beta$-function equations \cite{kkl}
is then given by functions $K$ which
satisfy the four-dimensional Laplace equation
\be
(\partial_u\partial_{\bar u}+\partial_v\partial_{\bar v})K=0.\label{laplace}
\ee
In addition, the dilaton field is simply given as
\be
2\Phi=\log K_{u\bar u}+{\rm constant}.\label{dilnfour}
\ee
 Eqs.(\ref{laplace}) and (\ref{dilnfour})
imply that $\delta c=0$ and these backgrounds are expected to have $N=4$
superconformal symmetry.  This obeservation
is consistent with the fact that eq.(\ref{laplace})
is the generalization of the hyper-K\"ahler
condition for spaces with antisymmetric
tensor field. The form of the dilaton field
  has the important consequence that the four-dimensional
metric in the Einstein frame is flat:
$
G_{\mu\nu}^{{\rm Einstein}}=e^{-2\Phi}G_{\mu\nu}^\sigma=\delta_{\mu
\nu}$.
 In fact,
the solutions
of the dilaton equation (\ref{dilnfour}) have a very close relation
to
the axionic instantons of \cite{rey}.
Specifically,  if we assume that the theory possesses two $U(1)$ isometries,
i.e. $\Phi=\Phi(u+\bar u,v+\bar v)$, equation (\ref{dilnfour})
implies
the following relation:
\be
{\rm d}\Phi=\pm {1\over 2}e^{-2\Phi}H^*.\label{selfdual}
\ee
This relation is nothing else than the self-duality
condition on the dilaton-axion field. Its solutions are
called axionic instantons.

Let us now
construct the dual spaces for the solutions
of the Laplace equation with   isometries.
We will perform a duality transformation  on the chiral $U$-field
replacing it by a twisted chiral field $\Psi$. The Legendre
transformed potential $\tilde K$ will only contain twisted
fields and will be therefore a true K\"ahler function
leading to a non-compact K\"ahler space without torsion.
Doing the Legendre transform we obtain the following line element
\be
{\rm d}s^2={1\over K_{uu}}({\rm d}z-K_{uv}{\rm d}v)({\rm d}{\bar
z}-K_{u\bar
v}{\rm d}{\bar v})-K_{v\bar v}{\rm d}v{\rm d}{\bar v}
\label{riflat}
\ee
where $K(u+\bar u ,v,\bar v)$ is the original quasi-K\"ahler
potential
and $z,\bar z$ are the dual coordinates defined via the Legendre
transform $z+\bar z =K_{u}$.
The Laplace equation implies that the determinant of the K\"ahler
metric
(\ref{riflat}) is constant so we obtain a Ricci flat K\"ahler
manifold.
The dual dilaton is consequently constant.
The metric (\ref{riflat}) describes a large
class of 4-d non-compact Calabi-Yau manifolds, which are also
hyper-K\"ahler.
The associated $\s$-models have N=4 superconformal symmetry and $\delta c=0$.
The manifolds have generically asymptotically flat regions as well
as curvature singularities.

Let us study now the  special case of
two isometries, i.e. $K(u+\bar u,v+\bar v)$.
If we paramertrize, $u=r_{1}+i\theta$, $v=r_{2}+i\phi$ then K is of
the form
$K(r_{1},r_{2})=iT(r_{1}+ir_{2})-i\bar T(r_{1}-ir_{2})$.
Introducing a new complex coordinate $z=r_{1}+ir_{2}$, we can write
the metric
(\ref{riflat}) in the following suggestive form
\be
{\rm d}s^{2}={Im{\rm T}\over 2}{\rm d}z{\rm d}{\bar z}+{2\over Im{\rm
T}}
({\rm d}\theta +{\rm T}{\rm d}\phi)({\rm d}\theta +{\bar {\rm T}}{\rm
d}\phi)\label{tor}
\ee
where T$(z)$ is an arbitrary meromorphic function.
Now the interpretation of the metric (\ref{tor}) is straightforward:
If we take $\theta,\phi$ to be angular variables, then they
para\-me\-tri\-ze
a 2-d torus, with modulus T$(z)$ which depends holomorphically
on the rest of the coordinates and conformal factor proportional
to $1/Im $T.

\section{Conclusions}
\setcounter{equation}{0}

We have examined some four-dimensional superstring backgrounds with
N=2 and N=4
superconformal symmetry (classical solutions to superstring theory).
We show that there exists a plethora of such theories with
non-trivial
metric, dilaton and antisymmetric tensor field.
It is a very interseting problem to find the exact $N=2$ and $N=4$
superconformal field theories which correspond to our general solutions
as it was already done for particular backgrounds in \cite{partic}.
Upon analytic continuation of the Euclidean solutions
we expect to obtain many cosmological solutions to superstring theory whose
spacetime
properties deserve further study.

\vskip .5cm

We would like to thank E. Kiritsis and C. Kounnas for a very pleasant
collaboration on the material presented in this report.
\noindent

\def\PL{Phys. Lett. }
\def\NP{Nucl. Phys. }
\def\PR{Phys. Rev. }

\end{document}